\documentclass[dvips]{acta}
\usepackage{lscape,epsfig}
\usepackage{amssymb}
\usepackage{amsmath}
\DeclareSymbolFont{ppa}{OT1}{ppl}{m}{it}
\DeclareMathSymbol{\vv}{\mathalpha}{ppa}{'166}

\newfont{\hb}{rphvb at 10pt}
\newfont{\hbo}{rphvbo at 10pt}
\newfont{\bitt}{rptmbi at 12pt}
\newfont{\bits}{rptmbi at 11pt}

\SetPages{0}{0}

\SetVol{62}{2012}

\begin{document}

\newcommand{\TabApp}[2]{\begin{center}\parbox[t]{#1}{\centerline{
  {\bf Appendix}}
  \vskip2mm
  \centerline{\small {\spaceskip 2pt plus 1pt minus 1pt T a b l e}
  \refstepcounter{table}\thetable}
  \vskip2mm
  \centerline{\footnotesize #2}}
  \vskip3mm
\end{center}}

\newcommand{\TabCapp}[2]{\begin{center}\parbox[t]{#1}{\centerline{
  \small {\spaceskip 2pt plus 1pt minus 1pt T a b l e}
  \refstepcounter{table}\thetable}
  \vskip2mm
  \centerline{\footnotesize #2}}
  \vskip3mm
\end{center}}

\newcommand{\TTabCap}[3]{\begin{center}\parbox[t]{#1}{\centerline{
  \small {\spaceskip 2pt plus 1pt minus 1pt T a b l e}
  \refstepcounter{table}\thetable}
  \vskip2mm
  \centerline{\footnotesize #2}
  \centerline{\footnotesize #3}}
  \vskip1mm
\end{center}}

\newcommand{\MakeTableApp}[4]{\begin{table}[p]\TabApp{#2}{#3}
  \begin{center} \TableFont \begin{tabular}{#1} #4
  \end{tabular}\end{center}\end{table}}

\newcommand{\MakeTableSepp}[4]{\begin{table}[p]\TabCapp{#2}{#3}
  \begin{center} \TableFont \begin{tabular}{#1} #4
  \end{tabular}\end{center}\end{table}}

\newcommand{\MakeTableee}[4]{\begin{table}[htb]\TabCapp{#2}{#3}
  \begin{center} \TableFont \begin{tabular}{#1} #4
  \end{tabular}\end{center}\end{table}}

\newcommand{\MakeTablee}[5]{\begin{table}[htb]\TTabCap{#2}{#3}{#4}
  \begin{center} \TableFont \begin{tabular}{#1} #5
  \end{tabular}\end{center}\end{table}}

\newfont{\bb}{ptmbi8t at 12pt}
\newfont{\bbb}{cmbxti10}
\newfont{\bbbb}{cmbxti10 at 9pt}
\newcommand{\uprule}{\rule{0pt}{2.5ex}}
\newcommand{\douprule}{\rule[-2ex]{0pt}{4.5ex}}
\newcommand{\dorule}{\rule[-2ex]{0pt}{2ex}}
\begin{Titlepage}
\Title{Puzzling Frequencies in First Overtone Cepheids}
\Author{W.\,A.~~ D~z~i~e~m~b~o~w~s~k~i}
{Warsaw University Observatory, Al.~Ujazdowskie~4, 00-478~Warsaw, Poland\\
Copernicus Astronomical Center, ul.~Bartycka~18, 00-716~Warsaw Poland\\
e-mail: wd@astrouw.edu.pl}
\vspace*{-3pt}
\Received{October 3, 2012}
\end{Titlepage}
\vspace*{-7pt}
\Abstract{The OGLE project led to discovery of earlier unknown forms of
multiperiodic pulsation in Cepheids. Often, the observed periods may be
explained in terms of simultaneous excitation of two or rarely three radial
modes. However, a secondary variability at about 0.6 of the dominant
period, detected in a number of the  first overtone (1O) pulsators inhabiting the
Magellanic Clouds, seems to require a different explanation. After
reviewing a possibility of explaining this signal in terms of radial and
nonradial modes, I find that only unstable modes that may reproduce the
observed period ratio are f-modes of high angular degrees ($\ell=42{-}50$).
I discuss in detail the driving effect behind the instability and show that
it is not the familiar opacity mechanism. Finally, I emphasize the main
difficulty of this explanation, which requires high intrinsic amplitudes
implying large broadening of spectral line.}{Galaxies: Magellanic Clouds
Stars: variables: Cepheids Stars: oscillations}
\vspace*{-3pt}

\Section{Introduction}
When a secondary periodicity is detected in a Cepheid, the initial
hypothesis is always that this is due to excitation of another radial
mode. There are some theoretical and observational arguments against
interpretation in terms of nonradial modes. None of the arguments is
compelling, especially if the amplitude of the secondary variability is
low. However, the interpretation in terms of radial mode is the most
attractive because only then the secondary periodicity yields immediately a
precise constraint on stellar parameters. The strength of information
contained in data on two radial mode periods is best seen in the impact of
the seminal Petersen's (1973) paper on stellar physics. The huge mass
discrepancy revealed in the diagram $P_S/P_L$ \vs $\log P_L$ known as the
Petersen diagram prompted the major stellar opacity revision. Such diagrams
are still commonly used to confront data with models.

In 1973 the list of double-mode Cepheid contained 8 objects. The number of
such objects has been considerably increased thanks to massive photometry
of stars in the Magellanic Clouds within the framework of the microlensing
projects: MACHO, OGLE, and EROS. In particular, the OGLE-III Catalog of
Cepheids in the LMC lists 61 objects pulsating simultaneously in
fundamental and first overtone modes (F/1O), and 203 in two consecutive
overtones (1O/2O). Similar numbers (F/1O -- 59, 1O/2O -- 215) of double
mode Cepheids are listed in the catalog for the SMC (Soszyñski \etal  2008). The
catalog contains also some rare objects with frequencies which may be
interpreted in terms of different combinations of radial modes.
Exceptionally valuable are objects where three radial modes are
identified. Data on the three periods allow for an independent
determination of the distance to Magellanic Clouds (Moskalik and
Dziembowski 2005). With a precisely determined distance, massive photometry
yields strong constraints on models of individual double-mode Cepheids.

In a number of Cepheids, the period of the secondary variability is
incompatible with the radial mode interpretation. First such cases were
discovered by Moskalik and Ko³aczkowski (2008, 2009) in their analysis of
OGLE-II data for the LMC. They found such periods in 42 objects with
dominant mode being first overtone and named them FO-$\nu$ Cepheids. In
most of the cases the secondary periods were close to the dominant modes
and were interpreted as an analogue of the Blazkho effect in RR~Lyr
stars. However, in 7 objects, the secondary periods were between 0.6 and
0.64 of the dominant (first overtone) period and required a different
interpretation. After rejecting the possibility that the secondary
periodicity is associated with another radial mode, the authors concluded
that the high-frequency signal is due to a nonradial modes but noted the
problem with driving such modes. The difficulties of this interpretation
were also briefly discussed by Dziembowski and Smolec (2009). It is
interesting that though the Cepheid masses (1--2~\MS) determined by
Petersen (1973) were by at least by a factor of 3 lower than accepted at his time,
still interpretations other than simultaneous excitation of F and 1O modes
were never considered in published papers.

The subsequent phase of the OGLE project led to the discovery of many more
Cepheids with the dominant first overtone and the period ratio in the
0.6--0.64 range. The OGLE-III Catalog provides data on 29 objects of this
type in the LMC and 139 in the SMC. They seem most puzzling Cepheids posing
interesting questions. First of all we would like to know what is the
nature of the secondary periodicity. We also would like to understand why
they are so much more frequent in the SMC than in the LMC and why there is
apparently no counterpart of this phenomenon among Cepheids pulsating in
the fundamental mode.

In Section~2 I review the data and reconsider the two-radial mode
interpretation, taking into account data on periods and reddening-free
magnitudes. On the model side, linear instability of considered modes is
set as the requirement. Interpretation of the secondary periodicity in
terms of nonradial mode excitation is studied in Section~3. Section~4 is
devoted to explanation of driving mechanism acting in various modes.
Visibility of high-degree modes which reproduce the whole range of puzzling
frequencies is discussed in Section~5.

\Section{The 1O/X Cepheids in the Magellanic Clouds}
\subsection{The LMC}
The difficulty of the two-radial mode interpretation for the 1O-X Cepheids
in the LMC is demonstrated in Fig.~1. The data shown there are from
Soszyñski \etal (2008) and they include in addition to the two periods, the
Wesenheit index, which is a reddening-free stellar luminosity measure
evaluated from mean {\it I} and {\it V} magnitudes with the standard
expression
$$W_I=I-1.55(V-I).\eqno(1)$$
\begin{figure}[htb]
\centerline{\includegraphics[width=12.5cm]{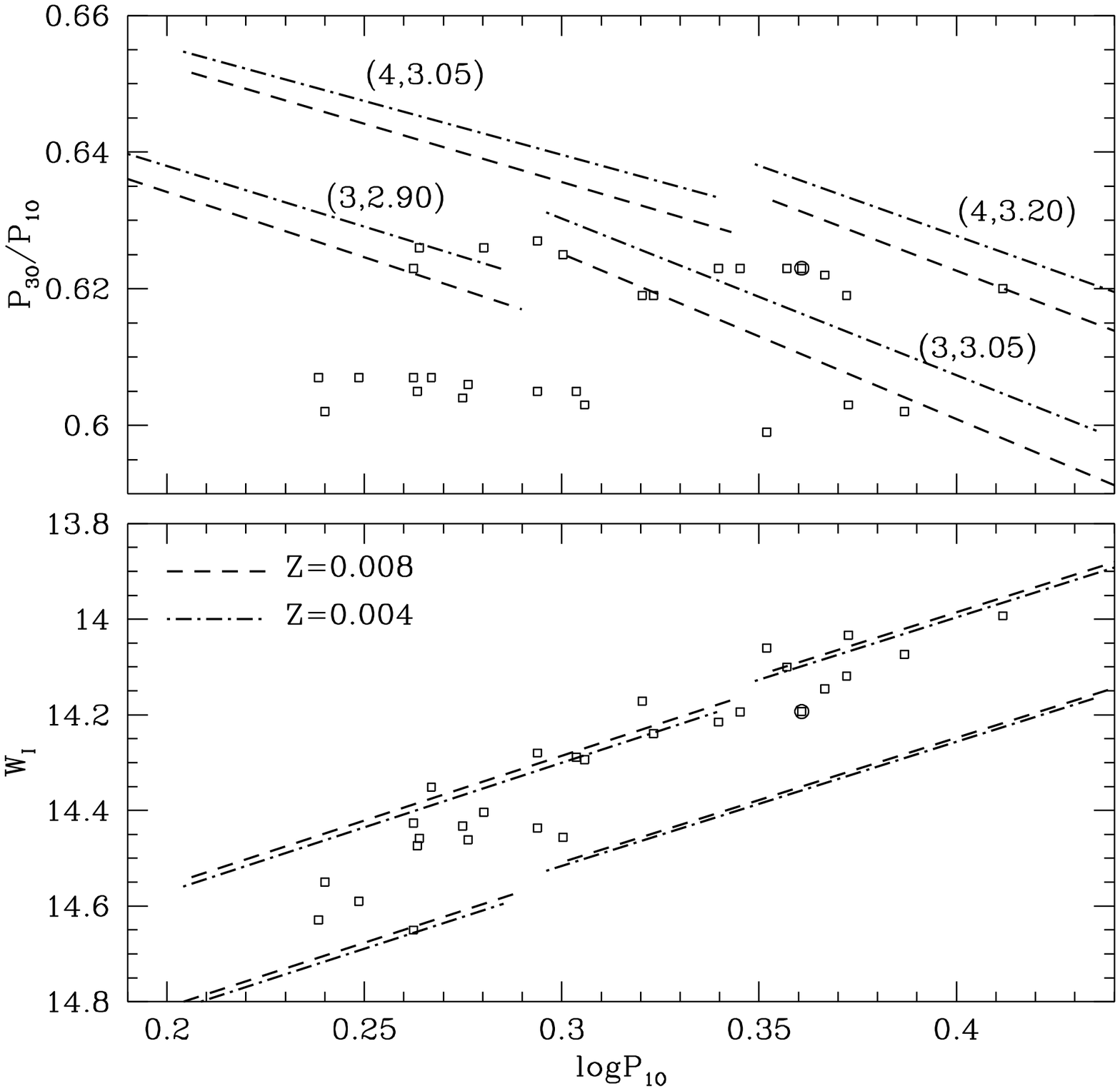}}
\vskip6pt
\FigCap{Comparison of the periods and the $W_I$ luminosity indices
for the 1O/X Cepheids in the LMC from the OGLE-III catalog with model
values. Observational data are shown with symbols. The encircled symbol in
{\it both panels} refers to the object where the fundamental radial mode is
excited along with the first overtone. The {\it upper panel} shows Petersen
diagram. The symbols are based on the measured $P_{1\rm O}$ and $P_{\rm X}$
periods. The calculated periods refer to the first and third overtones in
the envelope models with two indicated metal abundance parameters, $Z$,
shown with lines. Individual lines cover the range of $T_{\rm eff}$ of the
first overtone instability strip (see text for explanation), at specified
mass, $M$, and luminosity, $L$. The values of $M$ and $\log L$,
expressed in solar units, are given
above each pair of the lines. The {\it lower panel} shows the
period-luminosity relation employing the Wesenheit luminosity index, $W_I$,
(see Eq.~1).}
\end{figure}

The envelope models and radial pulsation periods were calculated with the
standard Warsaw codes (see, \eg Moskalik and Dziembowski 2005) for stellar
parameters leading to unstable 1O modes in the period range of the 1O/X
pulsators. Two sets of abundance parameters were adopted
$(Z,Y)=(0.008,0.256)$ and $(0.004,0.24)$, which are most often used for
modeling young stars in the LMC and SMC, respectively. For each selected
value of $Z$, $M$ (mass in solar units), and $L$ (luminosity in solar
units) two values of the effective temperatures were selected. The higher
one corresponding to the onset of the 1O instability defines blue edge
(BE). The one by 0.04 dex lower was adopted as the red edge (RE) of the
considered ranges of $T_{\rm eff}$. With our code we cannot determine the
red edge. The choice of 0.04 is in a crude agreement with results of
nonlinear modeling of Cepheid pulsation with time-dependent versions of the
 Mixing Length Theory (MLT, Feuchtinger \etal 2000) and the observed range of the $V-I$ color
(see Fig.~3).

\vspace*{9pt}
The MLT parameter $\alpha=1.5$ was adopted for this plot but the value of
$\alpha$ has only marginal influence on calculated periods. The growth
rates are more affected. With $\alpha=0.5$, BE and RE are shifted by about
$-0,015$ dex. An increase of $\alpha$ above 1.5 does extend the instability
range of the third overtone but even at $\alpha=2.5$ this modes remains
stable in the relevant range of stellar parameters. The ranges of
luminosity were chosen to cover the whole range of $\log P_{1\rm
O}$. Models used in this plots are realistic in the sense of physics but do
not follow from stellar evolution calculations. In fact, at the considered
$Z$ values no such calculation predicts entering the instability strip at
$M=3$ in the He burning phase. At $M=4$ some do but only models that
include overshooting reach $\log L>3$ in this phase.

\vspace*{9pt}
In the upper panel of Fig.~1, we may see that the objects split into two
nearly equinumerous groups with period ratios in the [0.599,0.607] and
[0.619,0.627] ranges. The objects occur in a rather narrow period range,
extending from 0.24 to 0.41 days. One of these objects (shown with the
encircled symbol) is an F/1O pulsator. The $P_{1\rm O}/P_{F\rm O}=0.714$
ratio at $\log P_F=0.506$ is well within the range of F/1O Cepheids in LMC
but the $A_{\rm F}/A_{\rm 1O}=0.04$ is exceptionally low. There is nothing
that significantly distinguishes this object from the rest in Fig.~1.

There is no doubt regarding 1O identification of the dominant mode. The
only radial mode that may be contemplated for association with $P_{\rm X}$
is 3O because for realistic stellar models $P_{2\rm O}/P_{1\rm O}\gtrsim
0.77$ and $P_{4\rm O}/P_{1\rm O}\lesssim0.55$ in the range of $P_{1\rm O}$
shown in Fig~1. At specified $P_{1\rm O}$ the $P_{3\rm O}/P_{1\rm O}$ ratio
depends both on $Z$ and $M$. To disentangle two effect we need data on
luminosity.

The model values of $W_I$ shown in the bottom panel were calculated with
the bolometric corrections and color indices are taken from Kurucz (2004)
assuming the 18.5~mag distance modulus to LMC. We see in the lower panel
that the dependence on $Z$ is very weak and that models of the same mass lie
nearly on a single line in the $\log P_{1\rm O}{-}W_I$ diagram. Thus, the
position in this diagram is a rather clean probe of stellar mass. We see
that the majority of the 1O/X Cepheids in the LMC have masses close
to~4. Only at short periods they are significantly lower but higher
than~3. At $M\approx4$ the plots in the upper panel leave the possibility
of the O3 interpretation for the secondary mode only for one object, which
is near $\log P_{1\rm O}=0.4$. Some of the data are reproduced by models
with $M\approx3$ but these are inconsistent with positions in the lower
panel. In many cases the fit requires $M$ significantly lower than~3. An
independent argument against the 3O identification for the secondary mode
comes from stability calculations which show that in all considered models
this mode is stable. Thus, even for the object at $\log P_{1\rm
O}\approx0.4$ such identification is excluded. In fact, it is the most
distant from the instability range.

\subsection{The SMC}
There are nearly five times as many 1O/X Cepheids in the SMC than in the
LMC and, as we may see in Fig.~2, the range of $\log P_{1\rm O}$ where the
objects occur is by a similar factor wider. The lack of objects with $\log
P_{1\rm O}<0.2$ in the LMC may be explained by much smaller total number of
the 1O Cepheids in this galaxy (see Fig.~9 of Soszyñski \etal 2010).
However, even at longer periods, the number of the 1O/X is by a factor of 2.4
larger though, beginning with $\log P_{1\rm O}=0.3$, the number of 1O
Cepheid in the LMC is greater than in the SMC. The differences in the
incidence of the 1O/X Cepheids between the two clouds are striking. In this
paper we focus mainly on the SMC data because much larger population
reveals the pattern which for the less numerous LMC objects may not be
visible.
\begin{figure}[htb]
\centerline{\includegraphics[width=12.5cm]{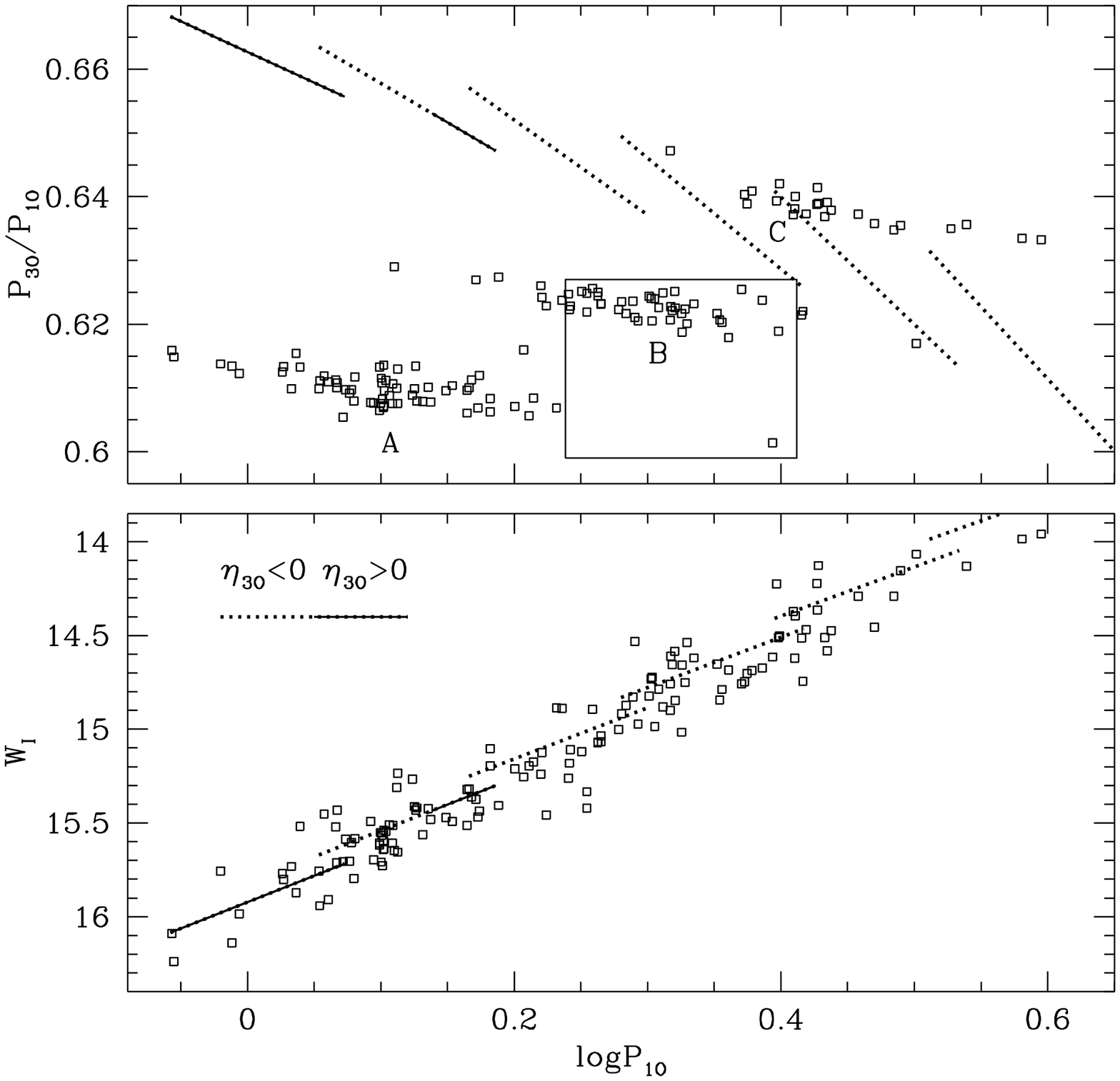}}
\vskip6pt
\FigCap{Comparison of the periods and the $W_I$ luminosity indexes
for the OGLE-III data (Soszyñski \etal 2010) for the 1O/X Cepheids in the
SMC with model values. Models used in this plot were calculated assuming
$Z=0.004$, and the $M{-}L$ relation given in Eq.~(2). Each of the six lines
extends over the O1 instability strip (as defined in the text) at fixed $L$
and $M$. In the part shown in the solid style also O3 modes are
unstable. The adopted range of $\log L$, is [2.7,3.45] and the resulting
range of $M$ is [3.2,5.17]. The rectangle in the {\it upper panel} shows
the range of the 1O/X Cepheids in the LMC.}
\end{figure}

\MakeTableee{cccccc}{12.5cm}{Three sequences of 1O Cepheids: mean parameters and
standard deviations}
{\hline\douprule
sequence& $P_{\rm X}/P_{1\rm O}$&$\log P_{1\rm O}$&$W_I$&$V-I$&No\\
\hline
 A&   $0.610\pm 0.003$&   $0.106\pm 0.071$&  $15.54\pm 0.27$&  $0.56\pm 0.06$&  65\\
 B&   $0.623\pm 0.002$&   $0.297\pm 0.066$&  $14.90\pm 0.30$&  $0.61\pm 0.05$&  50\\
 C&   $0.638\pm 0.003$&   $0.443\pm 0.065$&  $14.37\pm 0.25$&  $0.65\pm 0.04$&  24\\
\hline}

There are three well-detached sequences of 1O/X Cepheids concentrated
around $P_{\rm X}/P_{1\rm O}=0.610$, 0.623, and 0.638 marked in Fig.~2 as
A, B, and C, respectively. These ratios are correlated with periods but
there is some overlap. Within each of the sequence we see a slow decline of
the period ratio with $P_{1\rm O}$. The mean parameters of stars forming
the three sequences are listed in Table~1. Only the objects in sequence B
have their counterparts in the LMC. The objects in sequence A, except of
one, occur at much shorter periods than the LMC objects with similar period
ratios. There are no 1O/X Cepheids in the OGLE-III data for the LMC with
the period ratio as high as found in sequence C. It should be stressed,
however, that Moskalik and Ko³aczkowski (2009) in their analysis of the
OGLE-II data detected three such objects.

The color--magnitude and period--magnitude diagrams depicted in Fig.~3 show
that the 1O/X pulsators occur over nearly whole range of parameters of the
first overtone Cepheids in the SMC except that they are absent at shortest
periods. There are 320 first overtone Cepheids at $\log P_{1\rm O}<0.056$
but in none of them the puzzling high frequency signal has been detected
although the incidence of the 1O/X pulsators in the whole population is
about 8.5\%. This, in part, may be blamed to difficulty of detecting this
low amplitude signal in fainter objects but is unlikely to account for the
whole effect.
\begin{figure}[htb]
\centerline{\includegraphics[width=12cm]{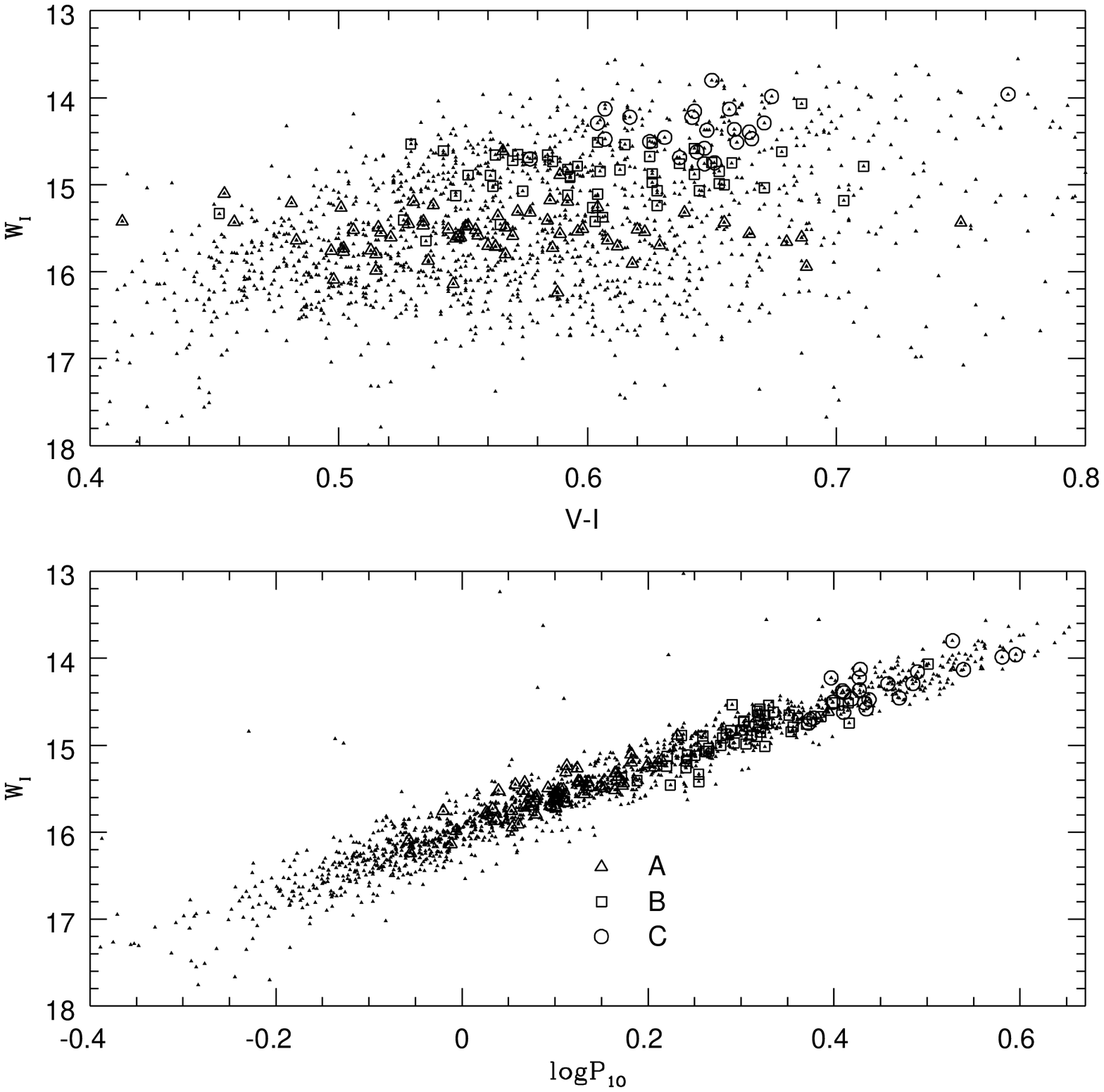}}
\vskip6pt
\FigCap{The 1O Cepheids in SMC from OGLE-III catalog in the CM
and PL diagram. The 1O/X objects from the three sequences are marked with
different symbols.}
\end{figure}

Models used for comparison with the data in Fig.~2 satisfy the mass--luminosity
relation
$$\log L=3.05+3.6(\log M-0.602),\eqno(2)$$
which is based on Girardi \etal (2000) evolutionary models of stars with
masses $M=4$ and 5 at $Z=0.004$. The tracks with $M=3$ do not reach the
instability strip in the helium burning phase. BaSTI (Pietrinferni \etal
2006) models also at $M=4$ and 5 do not rich the instability strip but the
maximum values of $L$ in this evolutionary phase is in a good agreement
with the relation given by Eq.~(2). Similar conclusion was reached by
Smolec (private communication), who used public domain code MESA (Paxton
\etal 2011). Apparently, new evolutionary codes do not describe well the
Cepheid phase. In the lower panel of Fig.~2, we may see that with the
adopted mass--luminosity relation the observational data in the $P{1\rm O}{-}
W_I$ plane are well reproduced. Interestingly, I found it also for the LMC
data. Models calculated with $Z=0.008$, which have lower $M$ at the same
$L$, produce lines running well above observational data. This suggests that the
occurrence of 1O/X pulsation requires low metal abundance and perhaps in
this way may explain the large difference in the incidence of such a
pulsation form between the two systems.

\subsection{Problems with Two Radial Modes Hypothesis}
The problem with reconciling  the secondary period with $P_{3\rm O}$ for
the SMC 1O/X Cepheid is different at short and long periods. In the first
case, it is the large discrepancy in the period ratios and in the second it
is the stability of the third overtone. It is not easy to find modification
of stellar structure resulting in the required change in the period
ratios. Any admissible changes in $Z$ cannot do it. A factor of  $\sim10$
increase suffices to get agreement for the sequence B but not for
sequence~A. The plots in Fig.~3 do no reveal any anomaly that would suggest
nonstandard models for the 1O/X Cepheids. The absence of these objects in
the short period range, which presumably is populated by stars crossing the
instability strip, may suggest that the occurrence of the secondary
periodicity is related to some property of the deep interior structure. One
conceivable feature affecting mainly frequencies of higher overtones could
be a composition jump at fractional mass above $M_r/M\approx0.6$ but it is
not expected in the relevant stellar models.

The stability problem may seem less severe because the determination of the
instability ranges suffers from the poor understanding of interaction
between convection and pulsation. However, especially for sequence C
objects, we are far from the period range where the third overtone is
unstable. This is seen not only in results of calculations but also in the
observational data. There is one object in the SMC at $\log P_{1\rm
O}\approx-0.65$, that is, well below Fig.~2 range where almost certainly
the third overtone is excited together with the first two overtones (see
Fig.~4 in Soszyñski \etal 2010). In the LMC there are three O1/O2/O3
Cepheids all located near $\log P_{1\rm O}=-0.25$ (see Fig.~2 in Soszyñski
\etal 2008). Again this is far from the 1O/X range.

Perhaps the correct interpretation of the 1O/X Cepheids requires going
beyond the linear pulsation theory. However, since the period ratios are
significantly different from the ratios of low integers, the only
possibility seems to be a modification of the mean stellar structure by the 1O
presence leading to a modification of 3O properties. This possibility
remains to be studied.

\Section{Nonradial Modes and the Puzzling Frequencies}
The problem with nonradial mode excitation in giants is the strong damping
in the radiative interior which is a consequence of huge values of the
Brunt-V\"ais\"al\"a frequency, hence very large radial wave
numbers. However as Osaki (1977) first showed there is a class of nonradial
modes that despite strong damping in the core are unstable owing to the
same driving effect as radial modes. The instability occurs if the modes
are sufficiently well trapped in the envelope and this is possible if the
angular degree is sufficiently high. In the 7~\MS\ Cepheid model, the
instability started at $\ell=4$ and the period close to 1O and at $\ell=6$
and the period close to the fundamental radial mode. Frequencies and growth
rates were determined for envelope models alone upon assuming the running
wave inner boundary condition. At the same time, I found (Dziembowski 1977)
similar unstable modes in RR~Lyr models. Formal derivation of the
inner boundary is given in that paper. There is one-to-one correspondence
between trapped nonradial and radial modes, thus following earlier habits,
we will use names f- and p$_n$-modes. In particular, the 1O mode
corresponds to p$_1$ nonradial modes, all having one node in the acoustic
cavity.

\begin{figure}[htb]
\centerline{\includegraphics[width=12.3cm]{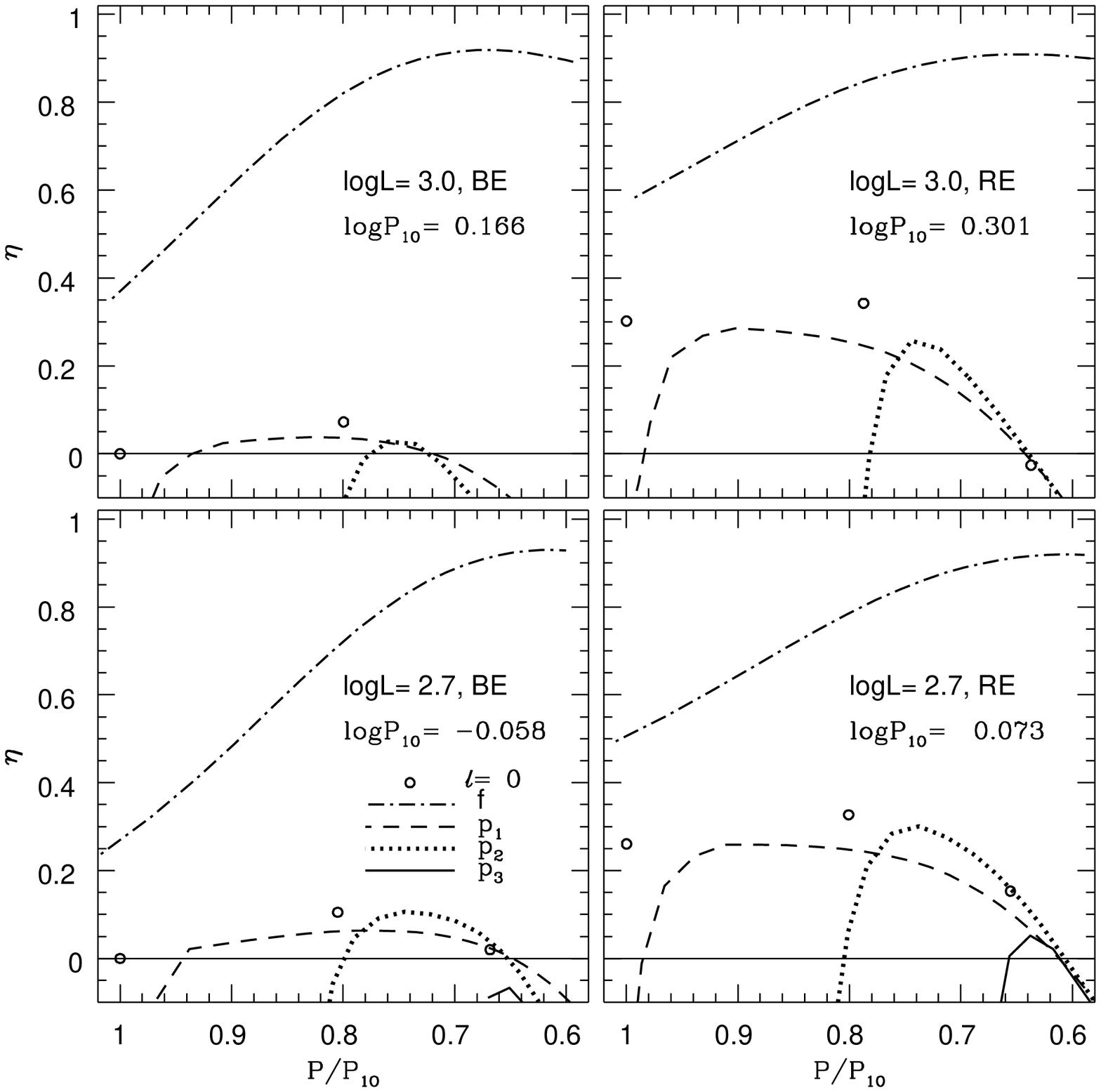}}
\vskip3pt
\FigCap{The relative driving rate for trapped acoustic modes and
high-degree f-modes in selected models ($M=3.2$ at $\log L=2.7$ and 3.87 at
$\log L=3.0$). In the $\log L=2.7$ BE (the 1O blue edge) model the
$\ell$-ranges of unstable p$_1$ and p$_2$ modes are [6,21] and [4,10],
respectively. In the cooler model (RE) at the same luminosity the
corresponding $\ell$-ranges are [5,22], and [3,12]. In this model there
are unstable p$_3$-modes at $\ell=3$, 4, and 5. In the $\log L=3$ BE,
model the ranges are [7,15], and [5,6] while in the $\log L=3$ RE model,
the ranges are [4,19], and [4,9]. In all models the f-mode shown have
$\ell$ ranging from 16 to 54 but the instability extends well beyond this
range.}
\end{figure}

Thirty years later, Mulet-Marquis \etal (2007) revisited the problem. They
conducted an extensive survey relying on complete stellar models and their
advanced tools for solving equations for nonradial oscillations. All
numerical results presented in the present paper were obtained with the
same code I developed over 35 years ago. A comparison with modern results
shows that this is adequate. We all share the same major uncertainty, that
is, the treatment of convection.

Fig.~5 depicts the relative driving rates
$$\eta=\frac{\int \dd^3{\pmb x} w}{\int \dd^3{\pmb x}|w|}\eqno(3)$$
where
$$w=\Im\left[\delta P\left(\frac{\delta\rho}{\rho}\right)^*\right]$$
is the local contribution to the work integral, for selected models of
Cepheid envelopes. The value of $\eta$ varies between $-1$ (no driving
zone) and and 1 (no damping zone). This parameter, which was introduced by
Stellingwerf (1978), is a better measure of the instability strength than
the usual growth rates and also better predictors for the nonlinear
development.

Only in the cooler (RE) model at the lowest $L$ do we find unstable low degree
modes. Unstable are p$_3$-modes at $\ell=3$, 4, 5. At $\ell=5$ the period
is significantly shorter than $P_{3\rm O}$ and the period ratio is not far
from that found in sequence A objects but the instability is marginal. At
lower degrees periods are very close to $P_{1\rm O}$. In any of the models
considered by us, not only those used in Fig.~4, the p-mode instability
range does not extend much beyond $P_{3\rm O}$ if this mode is
unstable. The f-modes are clearly different. They are strongly
unstable. The maximum of $\eta$ occurs in the observed range of $P_{\rm
X}/P_{1\rm O}$ but there is a difficulty following from high angular
degrees, which in this range are between $\ell=40$ and 50. Thus, large
pulsation amplitudes are needed to offset effect of cancellation in the mean
variability. We will return to this difficulty in Section~6. Let us first
try to explain why f-mode driving is so efficient.

\Section{The Driving Mechanism behind the f-Mode Instability}
That high-degree modes are unstable over wide range of stellar parameters
has been known since long time ago (Dziembowski 1977, Shibahshi and Osaki
1981). The authors of both papers noted the instability extends beyond the
blue edge for radial pulsation and that the hydrogen ionization is the main
site of the driving effect but they failed to explain the driving
mechanism. My aim in this section is to supply the explanation.

\begin{figure}[htb]
\centerline{\includegraphics[width=12.5cm]{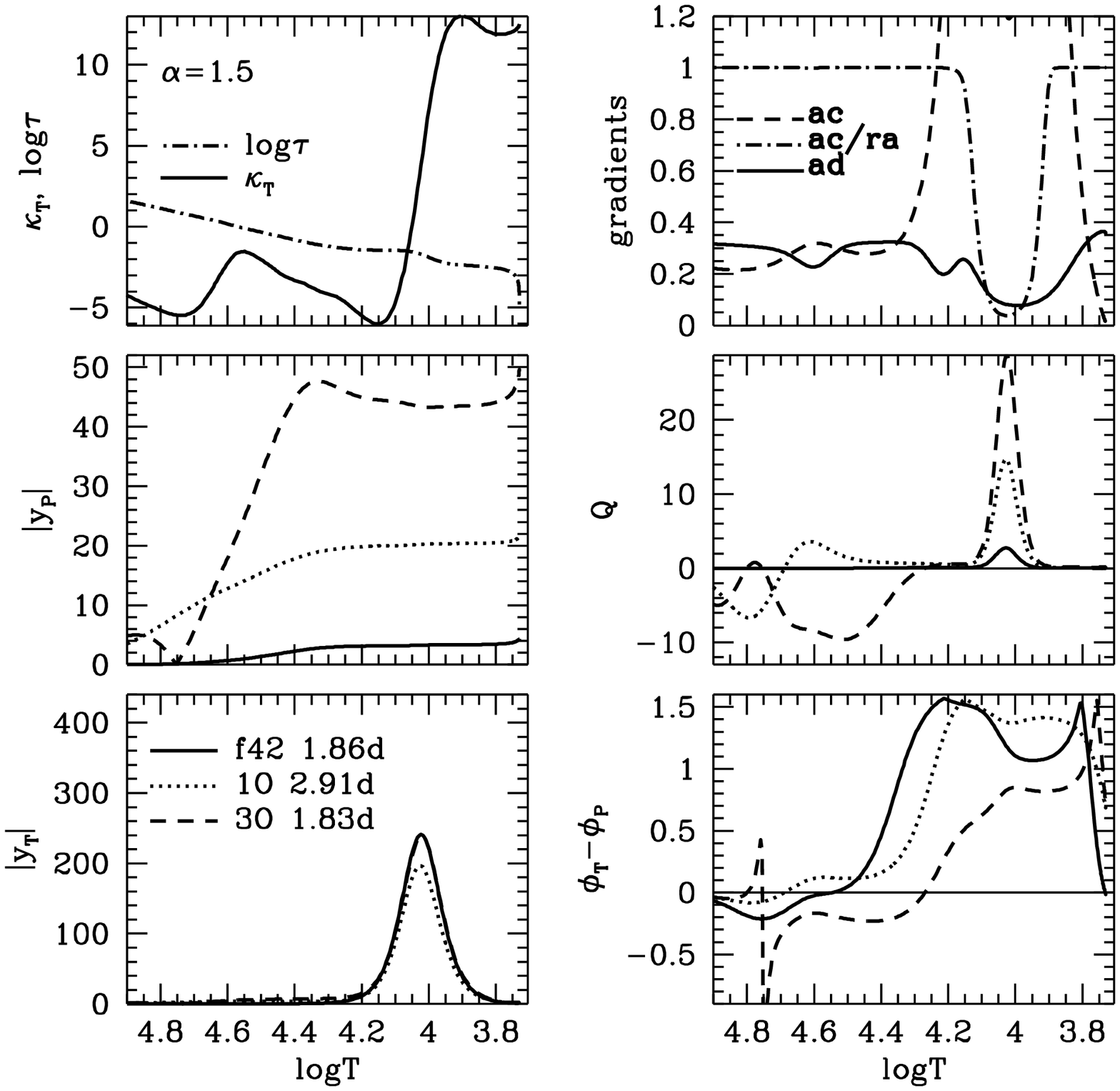}}
\vskip6pt
\FigCap{Plotted against the local temperature in outer layers are
various quantities relevant for understanding driving and damping effects
working in the three selected modes. The model is characterized by the
global parameters $M=4.69$, $\log L=3.3$, and $\log T_{\rm eff}=3.806$. It
was calculated assuming the mixing length parameter $\alpha=1.5$. The
selected modes are 1O and 3O radial modes and the $\ell=42$ f-mode
(f42). Their periods are listed in the {\it left bottom panel}. The {\it
top left panel} shows the opacity derivative
$\kappa_T=(\partial\log\kappa/\partial\log T)_P$ and the thermal time scale
in days, $\tau$ of the envelope above each point. The {\it upper right
panel} shows the actual temperature in the envelope $\dd\ln T/\dd\ln
P\equiv\nabla$, the ratio of actual to radiative
gradient,$\nabla/\nabla_{\rm rad}$, and the adiabatic gradient $\nabla_{\rm
ad}$. The {\it middle} and {\it panels on the left-hand side} show the
absolute values of the radial eigenfunctions for the pressure and
temperature perturbation, respectively. The {\it bottom right panel} shows
the phase difference between the two eigenfunctions. Plotted in the {\it
middle right panel} $Q$ (see Eq.~4) is the local contribution to driving
per unit of the independent variable.}
\end{figure}
\begin{figure}[htb]
\centerline{\includegraphics[width=12.7cm]{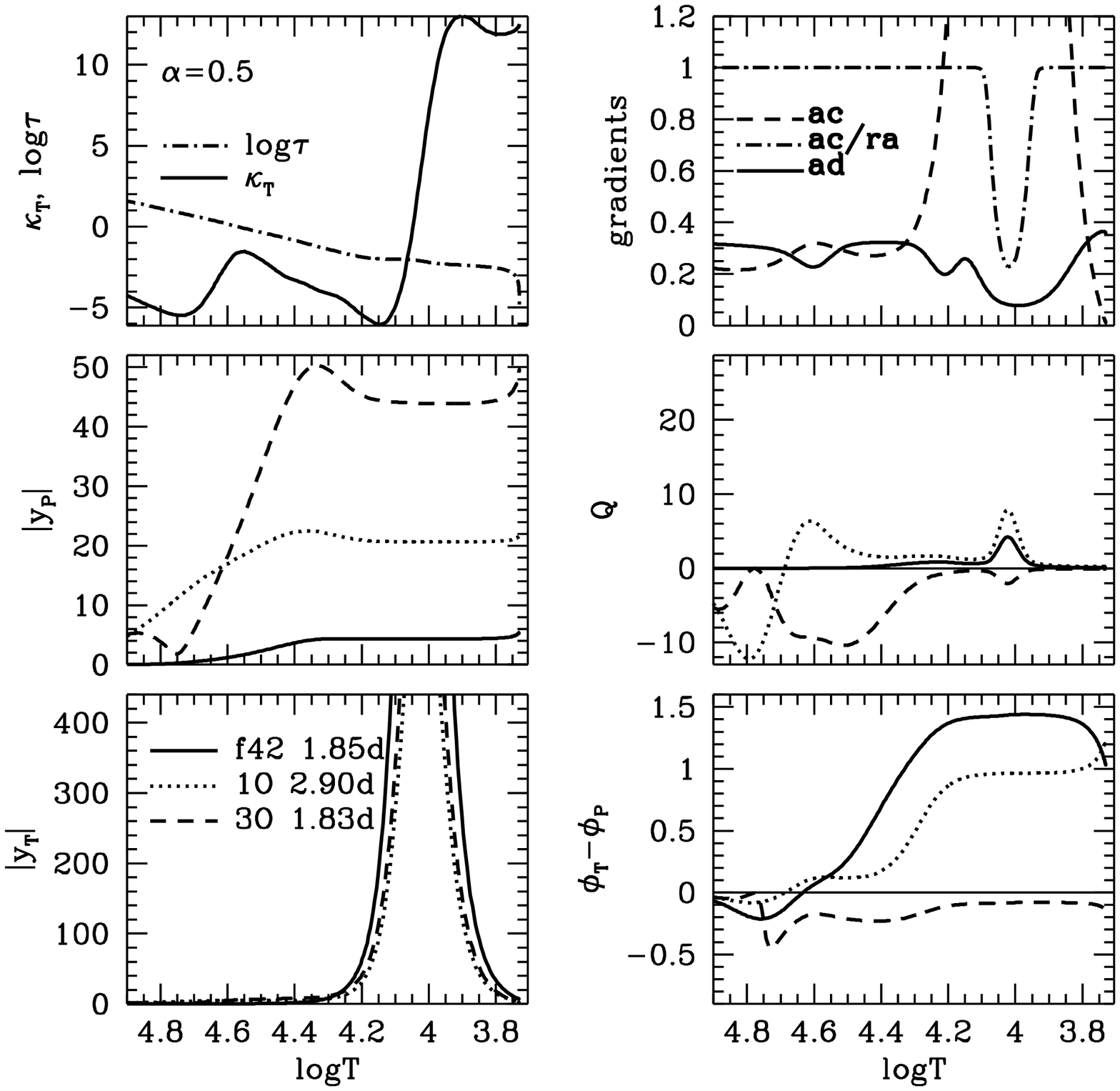}}
\vskip9pt
\FigCap{Same as Fig.~5 but for models calculated with $\alpha=0.5$
instead of 1.5.}
\end{figure}
Let us focus on two models with the same parameters at the surface but
differing in the efficiency of the convective transport. Plots in Fig.~5
and 6 refer to models calculated with $\alpha=1.5$ and 0.5,
respectively. The choice of $\alpha$ has a significant effect on
nonadiabatic mode properties. In particular, the relative driving rates for
the O1, O3, and $\ell=42$ f-mode (f42) are, respectively 0.29, $-0.65$,
and 0.87 at $\alpha=1.5$, while at $\alpha=0.5$ the corresponding rates are
0.04, $-0.99$, and 0.93. However, the huge difference in $\eta$, between O3
and f42 is preserved and this is true for all models considered in this
paper. This may look like a paradox. These two modes have nearly the same
periods; there is an additional energy loss in f42 resulting from the
horizontal radiative flux and yet this mode is driven in nearly the whole
outer part of the star where it is trapped. In Figs.~5 and 6 the driving
layers may be recognized as places where $Q>0$. This quantity is
proportional to $w$ (see Eq.~3) and it is given by
$$Q=Q_N\frac{Pr^3}{\nabla V}\left(-\frac{\partial\log\rho}{\partial\log
T}\right)_P\Im(y_P^*y_T)\eqno(4)$$
where $Q_N$ is the normalization constant, which is the same for all modes
in Figs.~5 and 6, $V=-{\dd\log P/\dd\log r}$, while $y_P(r)$ and $y_T(r)$
are radial eigenfunctions corresponding to $\delta P/P$ and $\delta
T/T$. The angular and time dependence of perturbations is assumed in the
form $Y_\ell^m\exp(-i\omega t)$. The standard normalization $y_r=1$ at the
surface was adopted for the radial eigenfunctions. In Figs.~5 and 6 in
addition to $Q$, plotted are $|y_P|$, $|y_T|$, and the phase difference
$\phi_T-\phi_P$, whose sign sets the sign of $Q$.

The classical opacity mechanism acts only in the HeII ionization zone and
only in 1O. Note rising $\kappa_T$ between $\log T=4.7$ and 4.55 and the
associated bump of $Q$ for this mode. For O3 this is the damping zone
because its amplitude varies rapidly and the radiative losses
dominate. Convection is present in this zone but carries a negligible
fraction of energy, even at $\alpha=1.5$. In the layers above the thermal
time-scale is much shorter than periods of the modes. Thus pulsation
becomes very nonadiabatic. The common feature of the modes is steep peak of
$|y_T|$ centered at $\log T\approx4.05$. Its height depends on mode and
$\alpha$ but neither on its localization nor its shape. In certain range
around its maximum, $|y_T|$ is roughly determined by the equation
$$\frac{\dd y_T}{\dd\log T}=2.3(\kappa_T-4) y_T+C\eqno(5)$$
where $C$, which includes all remaining terms from the expression for
perturbed radial component of the radiative flux, is regarded
constant. Associated with the $|y_T|$ peak is the peak of $Q$ at the same
location, except for 3O at $\alpha=0.5$ when we see there a deep. Always
higher $\alpha$ enhances driving effect in radial overtones and it is
opposite to f-modes, where higher $\alpha$ results in some reduction of
$\eta$. Apparently, in the latter case the effect of increased $|y_T|$
dominates over the effect of increased $\tau$. Regardless of the value of
$\alpha$ the phase difference does not change much in the layer around the
$|y_T|$ maximum. It growth takes place below, between $\log T=4.6$ and 4.2.
Thus, the driving mechanism actually works there and not in layers around
maximum of $Q$.

In order to identify the effect causing instability of the high-degree
f-modes, we transform  the expression for $Q$ (Eq.~4) using
$$\Im\left(\frac{\delta P^*}{P}\frac{\delta T}{T}\right)=
\Im\left(\frac{\delta P^*}{P}\frac{\delta S}{c_P}\right)=
\frac{1}{c_P}\Re\left[\frac{\delta P^*}{\omega
P}\delta({\pmb\nabla}{\pmb F})\right]$$
where ${\pmb F}$ denotes the total energy flux per unit area. Then, we
separate $Q$ into the part arising from the vertical ($Q_V$) and horizontal
($Q_H$) part of $\delta({\pmb\nabla}{\pmb F})$. Adopting the convective flux
freezing approximation in the form $\delta(r^2F_{\rm conv})=0$, we obtain
$$Q_V=\frac{Q_NL\nabla_{\rm ad}}{4\pi}\Re\left(\frac{y_P^*}{\omega}\frac{\dd
y_F}{\dd\ln T}\right),\eqno(6)$$
$y_F$ is the radial eigenfunction  corresponding to $\delta(4\pi r^2
F_{\rm rad})/L$, and
$$Q_H=-\frac{Q_NL\nabla_{\rm
ad}}{4\pi}\frac{\ell(\ell+1)}{\nabla\nabla_{\rm
rad}V^2}\Re\left(\frac{y_P^*}{\omega}(y_T+y_r\nabla V)\right).\eqno(7)$$
The two contributions and the net value of $Q$ in the f42 mode are plotted
in Fig.~7, where we may see that below $\log T=4.2$ the horizontal
energy losses lead to pulsation energy gain ($Q_H>0$). In order to
understand why is it so let us note that $(y_T+y_r\nabla V)$ is the radial
eigenfunction for the relative Eulerian perturbation of temperature,
$T'/T$. If $\omega$ is regarded real, which is not far from true, we have
$Q_H\propto\Re(\delta P*T')$. Thus, driving takes place if cooling, which
is proportional to $T'$, occurs in the low pressure phase and heating
occurs in the high pressure phase, just like in the well-known case of
semi-convective instability in the presence of the mean molecular weight
gradient. Long-time ago Souffrin and Spiegel (1967) found that the same
effect may cause instability of the gravity waves also in chemically
homogeneous layers. In the present situation, $Q_H>0$ because in the
considered range of $\log T$, the advection term $y_r\nabla V$ dominates
in $T'$  and the phase difference $\psi_r-\psi_P\approx\pi/2$ is largely
fixed by the inner boundary condition.
\begin{figure}[htb]
\centerline{\includegraphics[width=12.5cm]{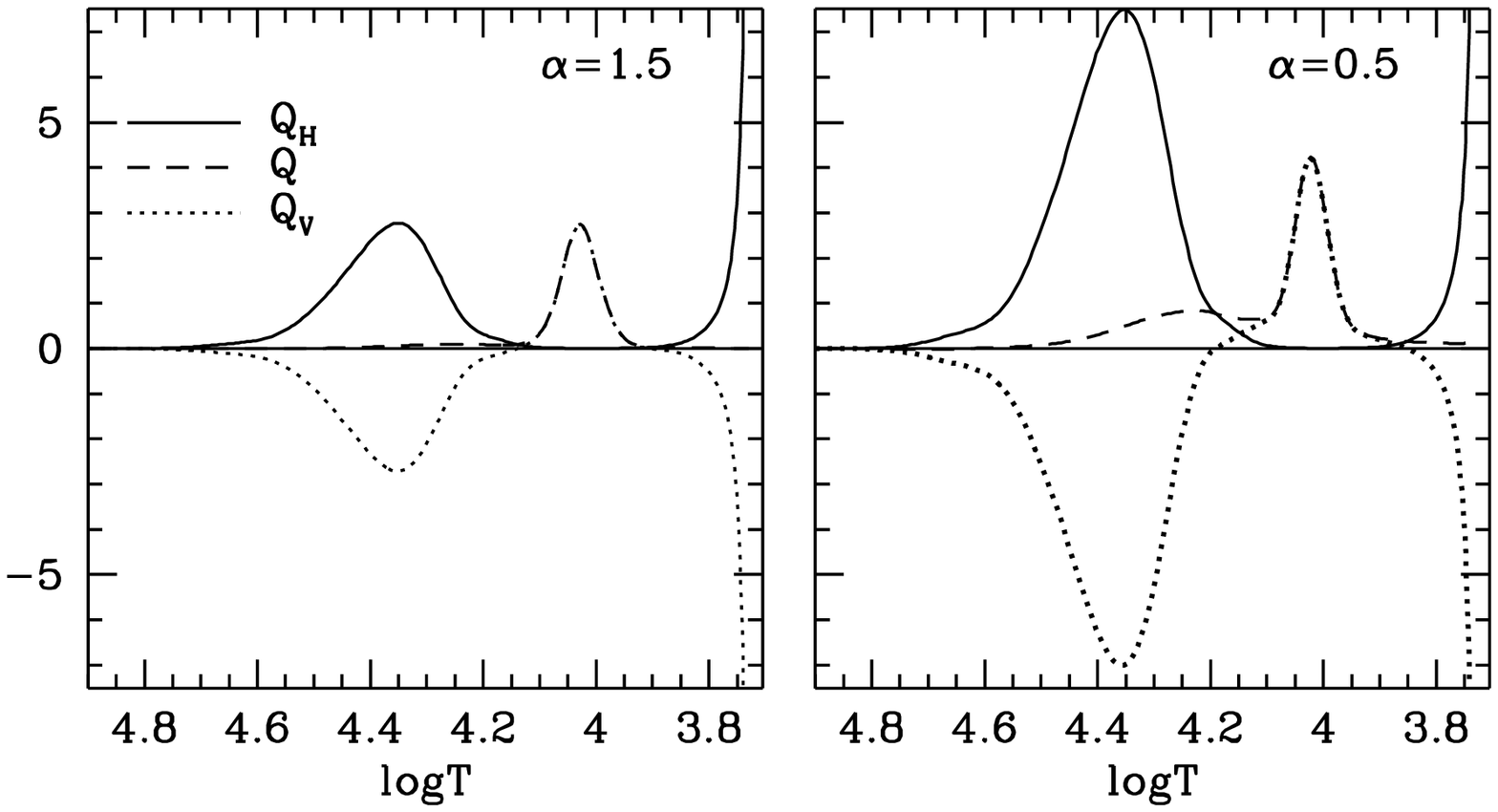}}
\vskip6pt
\FigCap{Values of $Q_V$ and $Q_H$ in models used in Figs.~1 and 2.}
\end{figure}

The pulsation energy gain from $Q_H>0$ is nearly canceled out by $Q_V<0$
because we are in the region where $\tau\ll P_{\rm X}$ and thus there is
a tendency toward the temporal thermal equilibrium. This changes near $\log
T=4.2$, where the steep rise of opacity inhibits the horizontal diffusion
and creates the $y_T$ bump, which enhances the vertical diffusion. Most of
the contribution to driving arises there but the essential phase lag
$\psi_T-\psi_P$ is created below.

\Section{Visibility of High Degree Modes}
We have noted in Section~3 that the f-modes that could explain the Petersen
diagram for the 1O/X Cepheid must have rather high degrees. The plots in
the upper panel of Fig.~8 suggest the $\ell=42$, 46, and 52 are excited in
sequence A, B, and C, respectively. The fact that only even degree modes
are seen is consistent with expectations because cancellation causes more
rapid decline of the disc-averaged amplitudes for odd-degree modes but why
multiples of four are preferred is difficult to explain. The p$_1$ modes in
the same frequency range have $\ell$'s by about a factor of 2 lower, hence they
might appear a better option for the identification. However, the
$\ell$-values best fitting for the three sequences are odd (19, 21, and 23)
hence are disfavored by the visibility argument. The arguments discussed
already in Section~4 also speaks against the p$_1$ option. In the RE
models, the unstable p$_1$ at $\ell=19$ occur only for $\log P_{1\rm
O}\lesssim0.3$, at $\ell=21$ for $\log P_{1\rm O}\lesssim0.1$, and at
$\ell=23$ for $\log P_{1\rm O}\lesssim0$, hence to the left of the great
majority of the data point. In the BE models the instability ranges are
still narrower. They are also narrower in models calculated with lower
$\alpha$.
\begin{figure}[htb]
\centerline{\includegraphics[width=12.5cm]{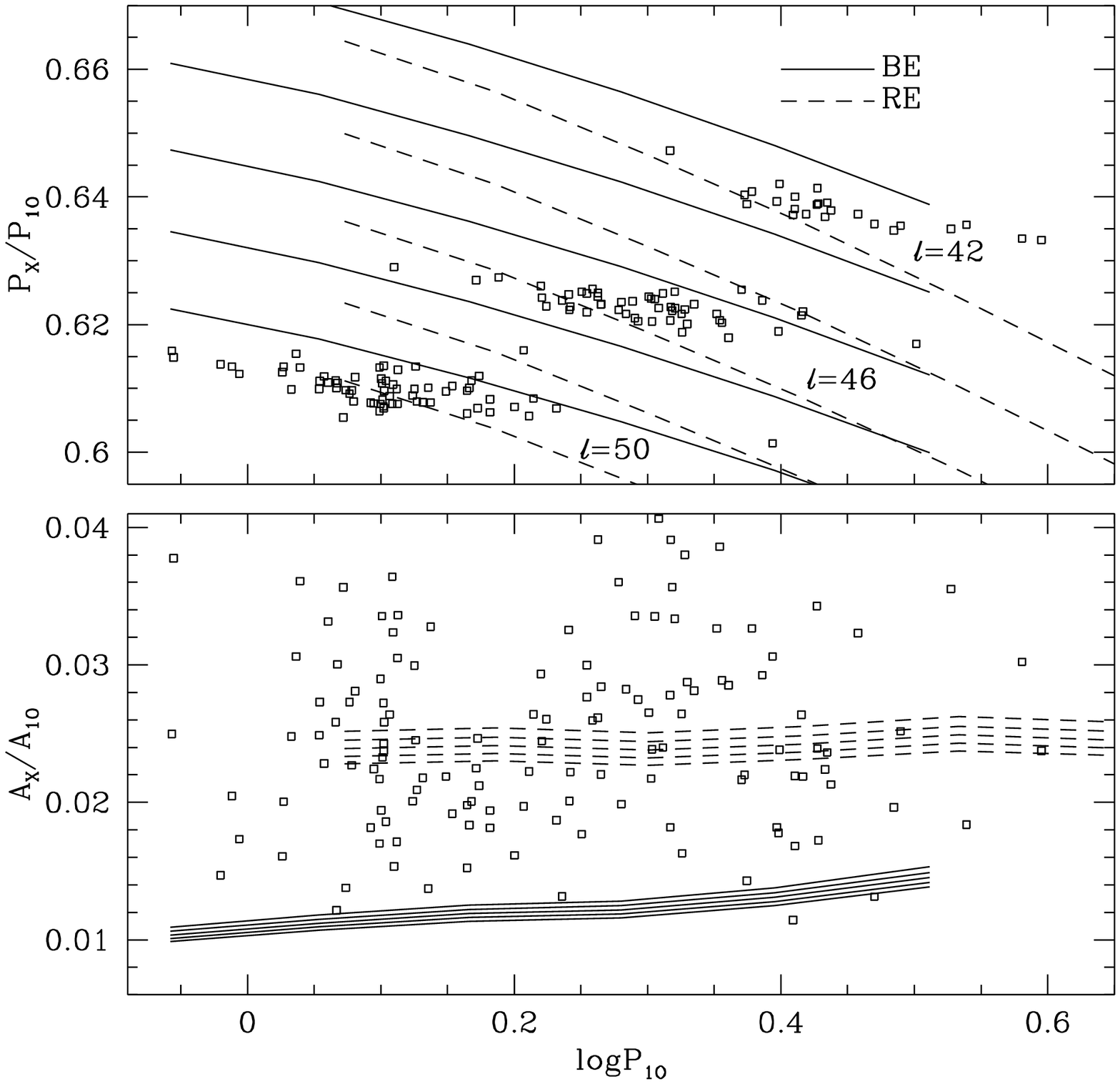}}
\vskip6pt
\FigCap{In the {\it upper panel}, the $P_{\rm X}/P_{1\rm O}$ period ratio
for the 1O-X Cepheids are compared with the evaluated ratios of
$P_{\ell}/P_{1\rm O}$ at even $\ell$'s for the same models as used in
Fig.~2. The labels showing mode degrees are put between the lines
corresponding to the red and blue edges of the instability strip. In the
{\it lower panel}, the observed amplitude ratios in the {\it I}-band are
compared with calculated amplitudes assuming the same intrinsic amplitudes
for the f-modes and the 1O mode.}
\end{figure}

The calculated amplitude ratios shown in the lower panel were obtained
assuming the same rms amplitudes of the photospheric radius variations,
$\overline{\delta R}_{\rm rms}$, for all modes. The numbers were obtained
with the use of parameters from Kurucz (2004) models and Claret (2000)
description of the limb-darkening law. The disc-averaged fluxes were
evaluated with help of the recurrence relation following Townsend
(2003). The observable amplitudes of the f-modes decrease slowly with
$\ell$ because in this range the dominant cause of the light changes is the
geometrical distortion. With the parabolic limb-darkening law, the
amplitude is proportional to $\ell^{-0.5}$ if $\ell$ is even. The decline
is also slow with the Claret version. Much larger ratios for the RE than BE
models follow primarily from much larger surface values of $y_F$ for the 1O
mode.

The comparison with observations suggests that $\overline{\delta R}_{\rm
rms}$ amplitudes of the f-modes would have to be of the same order as that
of the 1O mode to explain the 1O/X Cepheids. This is a very crude
conclusion because it is based on the linear relation between the flux and
radius amplitudes whereas at such intrinsic amplitudes we are then well
beyond validity of the linear approximation. We should also keep in mind
that calculated values refer to aspect-averaged amplitudes while the actual
amplitudes depend on aspect angle, $i$, and on the mode azimuthal order,
$m$, which we do not know. Certainly, the high intrinsic amplitudes of the
nonradial modes required for this interpretation is a serious
difficulty. However, with interpretation in terms of the p$_1$ modes, we
get the amplitude ratios lower only by some 50\%. Taking into account the
other difficulties of this hypothesis discussed earlier, we cannot regard
it as a viable alternative.

The difficulty of the interpretation in terms of the high-$\ell$
f-modes is best revealed with the estimate of the radial velocity
variations within stellar photosphere,
$$\vv_{\rm rad}(\theta,\phi,t)=\frac{A_\vv}{2}\Re
\left[\left(Y_\ell^m\cos i-\frac{1}{\ell}
\frac{\partial Y_\ell^m}{\partial\theta}
\sin i\right)\exp(-i\omega t)\right],$$
with $A_\vv=2\omega\overline{\delta R}_{\rm rms}$. This expression uses
$\omega^2=\ell GMR^{-3}$, which is approximately valid for these modes. The
value of ${\delta R}_{\rm rms}$ may be expressed in terms of the
aspect-average light amplitude. Keeping only the contribution from
geometrical distortion of the stellar surface, we get
$$\frac{A_\vv}{A_I}=0.047\frac{R}{P_{1\rm
O}}\left[\ell(\ell+1)\int_0^1h_I(\mu)\mu\dd\mu\right]^{-1} \quad
[{\rm km}/{\rm s}/{\rm mmag}]\eqno(8)$$
where the photospheric radius, $R$, is expressed in the solar units and
$P_{1\rm O}$ in days, while $h(\mu)$ describes the limb-darkening law. This
quantity, evaluated for the same models as used in Fig.~2, is shown in the
lower panel of Fig.~9. To estimate the local radial velocity we may use
data on light amplitude, which are plotted in the upper panel. The typical
velocity amplitudes range from 35 km/s for sequence C to 70~km/s for
sequence A. These are very large numbers. At so high $\ell$'s, the expected
signature is a nearly constant spectral line broadening. Such large effect
could not be overlooked.
\begin{figure}[htb]
\centerline{\includegraphics[width=12.3cm]{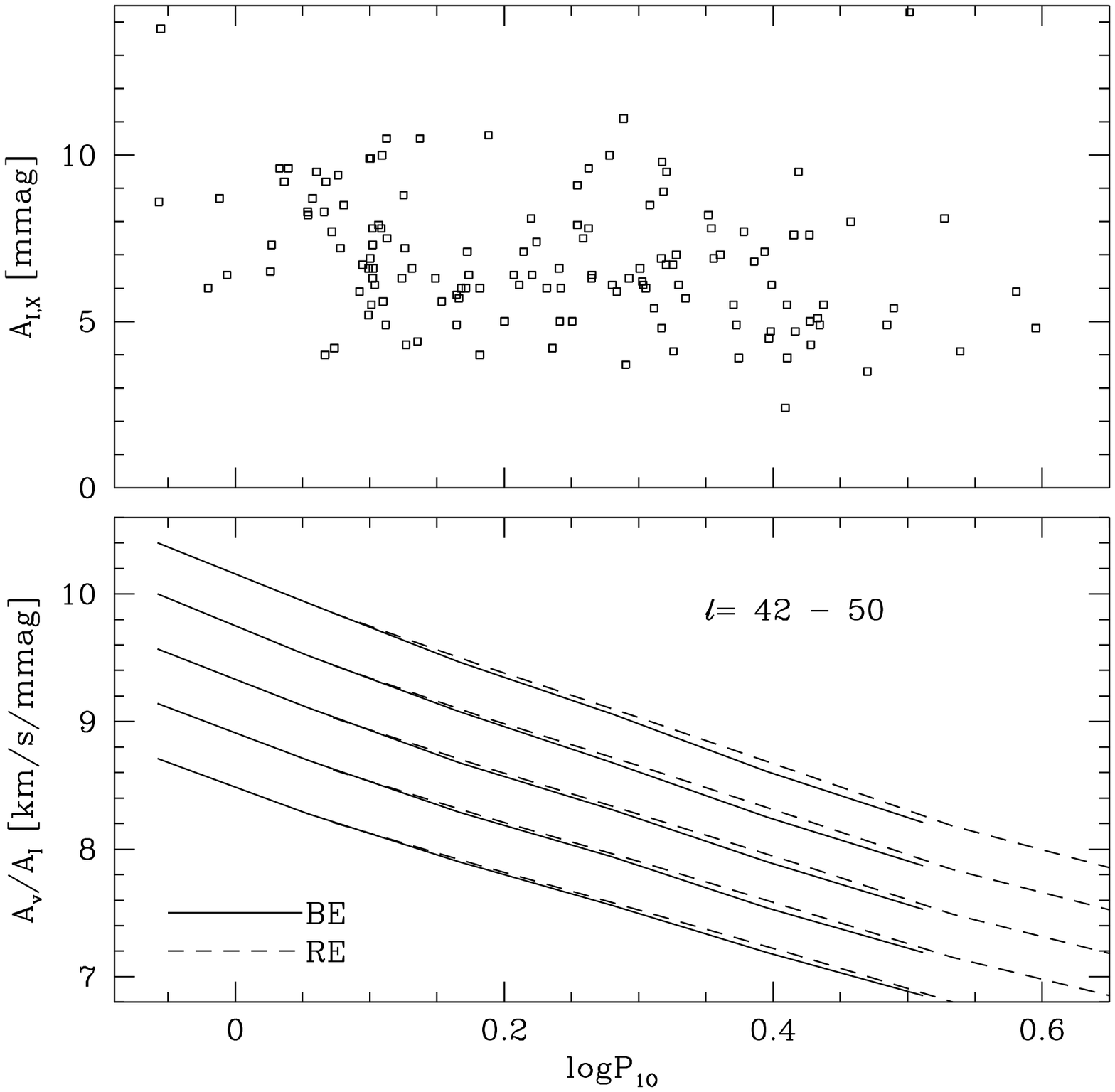}}
\FigCap{The {\it upper panel} shows the full light amplitude in the
{\it I}-band determined for the 1O-X Cepheids.The amplitude ratio plotted
in {\it lower panel} was evaluated with the use of Eq. (8) for same models
as used in Fig.~8.}
\end{figure}
Unusually broad line profiles ($\Delta\lambda/\lambda>10^{-4}$) were
observed in certain galactic Cepheids. Kovtyukh \etal(2003), report such a
phenomenon in four objects, of which three are certainly and one is
possibly first overtone pulsator. The authors noted significant changes in
the line profiles during pulsation cycle including moving bumps. Their
cautious interpretation is that low degree nonradial modes are excited in
these stars. Indeed, the observed patterns suggest rather \mbox{low $\ell$'s.}

The high degree mode instability is found over the whole luminosity range
of the classical instability strip and its blueward extension (Dziembowski
1977, Shibahashi and Osaki 1981). In all the cases the dominant
contribution to excitation arises in the hydrogen ionization zone. The
direct evidence that such modes are indeed seen in stars comes from
observations of the line profile variations in $\delta$~Sct stars. In
particular, Kennelly \etal (1998) report a detection of modes with $\ell$
up to 20 in $\tau$~Peg. Photometry alone may yield only circumstantial
evidences. For instance peaks at frequencies above the instability range
for p-modes could be interpreted as being due to high-$\ell$ f-modes if
they cannot be accounted for in terms of nonlinear effects of lower
frequency modes. This was the argument made by Daszyñska-Daszkiewicz \etal
(2006) regarding the high-frequency peaks in the $\delta$~Sct star
FG~Vir. However, we can never be sure that the inventory of the lower
frequency peaks is complete.

\vspace*{-9pt}
\Section{Summary and Discussion}
\vspace*{-5pt}
The secondary periodicity in the first overtone Cepheids at about 0.6 of
the dominant period presents a challenge to stellar pulsation theory. The
phenomenon is not rare. Soszyñski \etal (2010) found it in 8.5\% of the SMC
1O Cepheids. The incidence in the LMC is much lower, but still is seen
nearly 2.4\% of the objects, also the range of period is much narrower.
Objects exhibiting such secondary periodicity are called in this paper the
1O/X Cepheids. In the SMC, they form three detached sequences centered
around period ratios 0.61, 0.62, and 0.64 in the Petersen diagram. Except
for avoiding the shortest periods, there is nothing that distinguishes
these objects from the rest of the first overtone Cepheids.

One F/1O/X Cepheid was found in the LMC but the F-mode amplitude is
exceptionally low. No object that could be regarded as the F/X Cepheid has
yet been found. The phenomenon seems restricted to the first overtone
pulsators but not to Cepheids. Olech and Moskalik (2009) found the
secondary periodicity in seven RRc stars in $\omega$~Cen with the period
ratios between 0.608 and 0.622. Period ratios in the [0.612, 0.632] range
were found in three RRc stars observed with Kepler space telescope
(Moskalik \etal 2012). Earlier, a peak near $0.6P_{\rm1O}$ was found in the
RRd star AQ Leo by Gruberbauer \etal (2007).

Parameters of all 1O/X Cepheids in the Magellanic Clouds are well
constrained by Wesenheit index. Results of linear nonadiabatic calculations
for envelope models consistent with observed parameters seem to exclude
interpretation of the secondary periodicity in terms of a higher radial
overtone. The closest period is that of the third overtone but, in most
cases, it is far too long. In remaining cases the third overtone is stable. It
is unstable only at the shortest 1O periods, where the period ratio
discrepancy is the greatest. It is very difficult to imagine modification
in stellar models that could save the interpretation of the 1O/X Cepheids
as double radial mode pulsators.

A possibility that the secondary periodicity is due to a nonradial mode
trapped in the envelope was considered in Section~3. The conclusion was
that only high-degree f-modes remain unstable at the puzzling high
frequencies. The instability of such modes over wide ranges of frequencies
and stellar parameters has been known for years but has never been
satisfactorily explained. In Section~4, we could see that the crucial role
in the instability is played by the horizontal radiative transport, which
results in energy gain (loss) of the gas element in the high (low)
temperature phase. Nothing is known about nonlinear development of the
instability

The three sequences of the SMC 1O/X Cepheids corresponding to $P_{\rm
X}/P_{1\rm O}\approx0.64$, 0.62, and 0.61 may be associated with single
angular degrees $\ell=42$, 46, and 50, respectively. It is not surprising
that the degrees are even because amplitude reduction caused by
cancellation effect is stronger at odd $\ell$'s. No explanation was given
why the selected values are multiples of four. In fact, the greatest
difficulty of this interpretation are large values of $\ell$'s. Typical
amplitudes of variability in the {\it I}-band at the puzzling frequencies
are in the 4--10~mmag range. To reach such an amplitude against the
cancellation effect we must postulate the rms-values of $\delta R/R$
similar to that of the radial first overtone. Excitation of a high-$\ell$
mode with so large amplitude would manifests itself in a nearly constant,
large (35--70~km/s) spectral line broadening. No spectroscopic data are
available for any Cepheid known as a 1O/X object. However, overlooking the
low amplitude signal is more likely than the predicted large line
broadening. Still, even if the explanation in terms of f-modes looks
unrealistic, spectroscopic observations of the 1O/X Cepheids are most
desired because data on line profiles may yield a clue to the correct
explanation.

\Acknow{I am grateful to Pawe³ Moskalik for his helpful remarks after
reading preliminary version of this paper.}

\end{document}